# Resource Efficient Redundancy Using Quorum-based Cycle Routing in Optical Networks

**Cory J. Kleinheksel,** *Member, IEEE* **and Arun K. Somani,** *Fellow, IEEE*
*Electrical and Computer Engineering, Iowa State University, Ames, IA, USA*
*Tel: (515) 294 0442, Fax: (515) 294 9273, e-mail: {cklein, arun}@iastate.edu*

**ABSTRACT**
In this paper we propose a cycle redundancy technique that provides optical networks almost fault-tolerant point-to-point and multipoint-to-multipoint communications. The technique more importantly is shown to approximately halve the necessary light-trail resources in the network while maintaining the fault-tolerance and dependability expected from cycle-based routing.

For efficiency and distributed control, it is common in distributed systems and algorithms to group nodes into intersecting sets referred to as quorum sets. Optimal communication quorum sets forming optical cycles based on light-trails have been shown to flexibly and efficiently route both point-to-point and multipoint-to-multipoint traffic requests. Commonly cycle routing techniques will use pairs of cycles to achieve both routing and fault-tolerance, which uses substantial resources and creates the potential for underutilization. Instead, we intentionally utilize redundancy within the quorum cycles for fault-tolerance such that almost every point-to-point communication occurs in more than one cycle. The result is a set of cycles with 96.60 – 99.37% fault coverage, while using 42.9 – 47.18% fewer resources.

## 1. INTRODUCTION

We developed a novel method to deliver almost fault-tolerant capabilities of cycles in an optical network while significantly reducing the resource utilization when compared to the state-of-art techniques. Cycle-based routing can satisfy both dynamic point-to-point and multi-point optical communications. Cycles are created using quorums of nodes. Within a cycle, multicasts to all nodes in that cycle is possible. The quorum intersection property and the use of cyclic quorums sets provide all of the unicast capabilities. Exploiting the same properties we can achieve efficient broadcasts with $O(\sqrt{N})$ multicasts.

Optical networks are depended upon for high speed communications in distributed algorithms, as much as they are needed for the arbitrary point-to-point communications. Failures within a network are to be expected and can happen as much as every couple days. Protecting against these optical circuit faults is critical and there are many different approaches depending on the network needs and individual circumstances.

For efficiency and distributed control, it is common in distributed systems and algorithms to group nodes into intersecting sets referred to as quorum sets. Quorums sets for cycle-based routing to efficiently support arbitrary point-to-point and multi-point optical communication were first proposed in [1] with fault-tolerance analyzed in [2]. In this paper we apply the same established quorum set theory and add additional requirements to form suitable quorums for our optical network routing.

The rest of the paper is organized as follows. Sections 2, and 3 establish the network model, node communication, and path routing / fault-tolerance. In Section 4, we discuss our application of the distributed efficiency of the quorum sets to routing optical cycles. Lastly, Section 5 analyzes the performance of our redundant quorums set cycle routing techniques both without and with the presence of network link faults.

## 2. NETWORK MODEL

No two fiber-optic networks are the same. Some stretch hundreds of kilometers, while other networks are contained within buildings or rooms. These fiber-optic networks consist of several transmitters and receivers interconnected by fiber-optic cables. As you might expect, transmitters and receivers are typically found together and generically be called an optical node. The cables form the links (i.e. edges) between those nodes, which leads to a convenient model of a network in terms of a graph $G = (V, E)$. $V$ are the set of Nodes in the network and $E$ are the set of edges.

Edge $(a_i, a_j)$ is an fiber-optic link connecting nodes $a_i$ and $a_j$ in the network, where $a_i, a_j \in V$ and $(a_i, a_j) \in E$. It is a general assumption that the same set of optical wavelengths are available on all edges in $E$. The number of wavelengths available per optical fiber is dependent on the fiber-optic cables and the transmitter/receiver pairs.

## 3. LIGHT-TRAILS, CYCLE ROUTING, AND FAULT-TOLERANCE

Lightpaths were a critical building block in the first optical communications, but required significant traffic engineering and aggregation to support point-to-point communication. Light-trails were proposed in [3] as a solution to the challenges facing lightpaths and could be built using commercial off-the-shelf technology.

Light-trails enable fast, dynamic creation of an unidirectional optical communication channel allowing for receive and transmit access to all connected nodes. Communications from an upstream node to any number of downstream nodes can be scheduled on the shared light-trail. Being all optical avoids any energy losses and time delays associated with unnecessary Optical-to-Electrical-to-Optical (O/E/O) conversions at intermediate hops.

Figure 1 is simply a light-trail where the start and end node is the same node, referred to as the *hub node*. The hub node has its optical shutters in the *off* state, while intermediate nodes have their optical shutters in the *on* state. This effectively isolates an optical signal to a specific light-trail and allows for reuse of optical wavelength(s) elsewhere in the network.

Failures within an optical network are to be expected. P-cycle protection of unicast and multicast traffic networks requires preconfiguration and the offline nature allows for the efficient cycles to be selected [4]. The Optimized Collapsed Rings (OCR) single link protection heuristic was developed to address the heterogeneous, part multicast / part unicast, nature of WDM traffic [5]. ECBRA uses bidirectional cycles for fault-tolerance and outperforms the OCR heuristic [6] with only a $O(|E||C|^3)$ computational complexity.

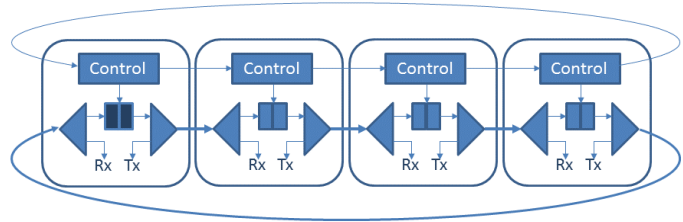

Figure 1. Cycle formed using the light-trail architecture.

## 4. QUORUMS

In distributed communication and algorithms, coordination, mutual exclusion, and consensus implementations have grouped N nodes into small sets call quorums. This organization of nodes can minimize communications in operations like negotiating access to a global resource.

### 4.1 Defining quorums set

A set of subsets (quorums) covering N entities and all quorums having non-empty intersections is a Quorums set.

$$S_1 = \{a_i, \ldots\}, i \in 1, 2, \ldots, N \qquad (1)$$
$$\bigcup_{i=1}^{N} S_i = \{a_1, \ldots, a_N\} \qquad (2)$$
$$S_i \cap S_j \neq \emptyset, \forall\, i, j \in 1, 2, \ldots, N \qquad (3)$$

The lower bounds for the maximum individual quorum size (i.e. $|S_i|$) in a minimum set is K, where Equation 4 holds and $(K-1)$ is a power of a prime [7]. Additionally it is desirable that each quorum $S_i$ in the quorum set be of equal size, Equation 5, such that there is equal work and it is desirable that each node be present in the same number of quorums, Equation 6, such that there is equal responsibility.

$$N \leq K(K-1) + 1, \qquad (4)$$
$$|S_i| = K, \forall\, i \in 1, 2, \ldots, N, \qquad (5)$$
$$a_i \text{ is contained in K } S_j's, \forall\, i \in 1, 2, \ldots, N \qquad (6)$$

Cyclic quorums adhere to these properties [8]. Cyclic quorums are unique in that once the first quorum is defined the remaining quorums can be generated via incrementing. For simplicity assume that $a_1 \in S_1$ without loss of generality (any one-to-one re-mapping of entity ids can result in this assumption.)

$$S_1 = \{a_1, \ldots, a_K\} \qquad (7)$$
$$S_i = \{a_{1+(i-1)}, \ldots, a_{K+(i-1)}\} \text{ modulo } (N+1) \qquad (8)$$

### 4.2 Finding redundant cyclic quorums sets

When K is minimal, every pair of entities $(a_i, a_j)$ would naturally occur together within a quorum at least once. Optical networking requires all directional point-to-point pairs to exist, i.e. both pairs $(a_i, a_j)$ and $(a_j, a_i)$. The existing state of art addresses this by pairing each cycle with the same cycle with its direction reversed.

In this paper, we are proposing moving the redundancy from the paired cycles and putting the redundancy in the quorums themselves. The key change is that every pair $(a_i, a_j)$ would must now occur together within two distinct quorums. Originally a minimal quorums set would have had $O(NK^2)$ pairs of entities, and now our proposal doubles the number of these pairs to $O(2NK^2)$. The number of quorum sets, N, is constant, which forces the new pairs to be generated from an increase in the minimal K of Equation 4 to a new redundant minimal quorum set size, $\widehat{K}$. Solving for $\widehat{K}$ in $O(N\widehat{K}^2) = O(2NK^2)$, reveals that the quorum size growth is not linear, instead $\widehat{K} \cong \sqrt{2}K$.

Using this reduced growth rate to our advantage, many point-to-point pairs will be created without substantially increasing the resources used. And this growth rate is far slower than simply duplicating a cycle.

To the best of our knowledge, no efficient algorithm is known to find quorums of minimum size, particularly those that have the additional requirement that entity pairs appear twice within the quorums set solution. Hence hundreds of compute hours were dedicated to exhaustively searching for minimal $\hat{K} \cong \sqrt{2}K$ cyclical quorums of size $\hat{K} \cong \sqrt{2}K$. The resulting redundant quorums were analyzed for their efficacy in optical quorum-based networking in the next section.

## 5. NETWORK ANALYSIS

To examine our proposed redundant quorum we used four common networks found in the prior art and an implementation of the ECBRA heuristic (proposed in [6]) to perform the cycle routing. ECBRA is sensitive to node and edge numbering that a total of 1000 variations on the inputs were considered, each being a one-to-one mapping with the respective network. For simulation of prior art in [1, 2], we used the $N = 4,...,111$ optimal cyclic quorums from [8]. Redundant cyclic quorums were found using the techniques described in Section 4.2.

### 5.1 Fault-free operational analysis

It is expected that a majority of the time the optical network will be operating without faults. It is important that the performance and resource utilization during this period be analyzed.

Table 1 shows the dramatic 42.9 – 47.18% resource reduction by employing our redundant quorums set technique to cycle routing. This reduction represents the potential for lower capital costs in terms of physical transmitters and receivers needed and the potential for more (wavelength) resource availability within the network. The state of art technique from literature uses paired cycles to form the quorum-based cycles and the mean network links used is shown in the Table 1, column three with a 95% confidence interval (CI.) Our redundant quorum technique uses far fewer links (shown in column four and five.) For reference our redundant quorum technique requires only a few more links than the state of art prior to their required pairing of cycles (column two.)

*Table 1. Mean links used by quorum cycle solutions (95% CI)*

| Network | Single Cycles | Paired Cycles | Redundant Cycles | Reduction |
|---|---|---|---|---|
| NSFNet – 14 Nodes, 22 Links | 124.48 – 124.94 | 248.96 – 249.88 | 135.10 – 135.46 | 45.76% |
| Arpanet – 20 Nodes, 31 Links | 255.02 – 255.62 | 510.03 – 511.03 | 269.44 – 269.93 | 47.18% |
| American Backbone – 24 Nodes, 43 Links | 320.99 – 321.68 | 641.98 – 643.35 | 359.44 – 360.10 | 44.02% |
| Chinese Backbone – 54 Nodes, 103 Links | 1336.80 – 1339.10 | 2673.60 – 2678.21 | 1526.70 – 1528.95 | 42.90% |

The authors in [1] considered only paired light-trails, which were needed in order to form all of the point-to-point communication node pairs with minimum sized quorum cycles. This work considers utilizing intentionally formed redundant node pairs within the quorum routing to reduce the resources used as a potential trade-off to a small cost to network performance. To measure this cost, metrics of missing node pairs is used.

Table 2 shows two important results. First, the dramatic reduction in resource utilization came at a tradeoff of a few missing communication pairs (<1%, column 5.) The paired cycles (column 4) used significantly more resources and didn't miss any pairs. Secondly, compared to single cycles (prior to pairing,) our redundant cycles perform 2x-4x better every time. As seen in Table 1, this performance improvement came at a much smaller cost than the state of art approach, where resource usage is doubled.

*Table 2. Mean missing node pairs by quorum cycle solutions (95% CI)*

| Network | Pairs | Single Cycles | | Paired Cycles | | Redundant Cycles | |
|---|---|---|---|---|---|---|---|
| | | # Pairs | % Pairs | # | % | # Pairs | % Pairs |
| NSFNet | 182 | 3.77 – 4.02 | 2.07% – 2.21% | 0.00 | 0.00% | 1.47 – 1.64 | 0.81% – 0.90% |
| Arpanet | 380 | 3.38 – 3.66 | 0.89% – 0.96% | 0.00 | 0.00% | 1.25 – 1.42 | 0.33% – 0.37% |
| American Backbone | 552 | 10.73 – 11.22 | 1.94% – 2.03% | 0.00 | 0.00% | 2.74 – 2.98 | 0.50% – 0.54% |
| Chinese Backbone | 2862 | 32.83 – 33.91 | 1.15% – 1.18% | 0.00 | 0.00% | 7.45 – 7.89 | 0.26% – 0.28% |

The quorums set method guarantees that all of the pairs exist (Section 5.2.) It is the limitations of an unidirectional optical light-trails with its required one optical shutter in the *off* state per cycle that has caused the missing pairs and the potential need for additional compensation steps. Compensation possible using an off-the-shelf solution of an additional routing step involving an Optical-to-Electrical-to-Optical (O/E/O) conversion and resending by a hub node.

## 5.2 Fault-tolerant operational analysis

Optical networks are highly depended upon. The fault-tolerance aspect of this route design is critical. Maintaining the ability to serve all dynamic point-to-point traffic requests despite fault is important. We assume a single fiber link failure. While a simple model, it does cover most real single fault scenarios. This occurs when a link is broken, like planned maintenance or the accidental severing during land excavation. Each modeled node has transmitters and receivers that can fail too. Short of a natural disaster, devices will likely fail independently of one another. When a transmitter/receiver pair fails within a modeled node, the effect on the global network is similar to that link failing. Modeling as a single edge failure, while not an exact fault mapping, is an appropriate abstraction.

To model the fault, we simulate the failure of each edge, $e \in E$, in the network model, $G = (V, E)$. We then examine the network's ability to serve all potential point-to-point requests by counting pairs of nodes that would be able to communicate and conversely those pairs that are unable to communicate.

Our simulation results showed our redundant quorum-based cycle technique had 96.60 – 99.37% fault coverage in the four networks tested. In Table 3, we compare the state-of-art paired cycle approach with our technique that uses significantly fewer resources. With single edge failures, the paired cycles had a mean missing communication pair rate of less than 3 pairs or less than 0.53% across all networks (95% CI.) Our redundant quorum cycles technique in column 4 could not reach that level of coverage, but did deliver an acceptable almost fault-tolerant technique at most a 3.40% mean missing pair rate (95% CI.)

*Table 3. Mean missing node pairs during single edge network failures (95% CI)*

| Network Name | Pairs | Paired Cycles | | Redundant Cycles | |
|---|---|---|---|---|---|
| NSFNet | 182 | 0.92 – 0.96 | 0.50% – 0.53% | 6.09 – 6.19 | 3.34% – 3.40% |
| Arpanet | 380 | 0.74 – 0.78 | 0.20% – 0.20% | 6.75 – 6.87 | 1.78% – 1.81% |
| American Backbone | 552 | 2.02 – 2.08 | 0.37% – 0.38% | 9.92 – 10.05 | 1.80% – 1.82% |
| Chinese Backbone | 2862 | 2.75 – 2.80 | 0.10% – 0.10% | 18.08 – 18.21 | 0.63% – 0.64% |

## 6. Conclusions

In this paper we proposed and evaluated a quorums set cycle redundancy technique that provides optical networks almost fault-tolerant communications. Quorums sets were chosen such that all network communication pairs appeared twice within a routed cycle set. We intentionally utilized this redundancy within the quorum cycles for fault-tolerance and reduction in resource usage. The result was a set of almost fault-tolerant cycles that used significantly fewer resources (42.9 – 47.18% fewer,) while at the same time maintaining a high degree of necessary fault-tolerance with 96.60 – 99.37% fault coverage. In future work, we are examining ways to improve both the fault and fault-free performance further, while maintaining the significant resource usage reductions.

## ACKNOWLEDGEMENTS

Research funded in part by NSF Graduate Research Fellowship Program, IBM Ph.D. Fellowship Program, Symbi GK-12 Fellowship at Iowa State University, and the Virginia and Phillip Sproul Professorship at Iowa State University. Any opinions, findings, and conclusions or recommendations expressed in this material are those of the author(s) and do not reflect the views of the funding agencies.

## REFERENCES

[1] A. K. Somani and D. Lastine, "Optical Paths Supporting Quorums for Efficient Communication," in *Optical Communications and Networks (ICOCN), 2014 13th International Conference on*, 2014.
[2] C. Kleinheksel and A. K. Somani, "Optical Quorum Cycles for Efficient Communication," *Photonic Network Communications,* 2015.
[3] I. Chlamtac and A. Gumaste, "Light-trails: A solution to IP centric communication in the optical domain," in *Quality of Service in Multiservice IP Networks*, Springer, 2003, pp. 634-644.
[4] F. Zhang, W.-D. Zhong and Y. Jin, "Optimizations of-Cycle-Based Protection of Optical Multicast Session," *Lightwave Technology, Journal of,* vol. 26, no. 19, pp. 3298-3306, 2008.
[5] A. Khalil, A. Hadjiantonis, G. Ellinas and M. Ali, "Pre-planned multicast protection approaches in WDM mesh networks," in *Optical Communication, 31st European Conference on*, 2005.
[6] A. K. Somani, D. Lastine and S. Sankaran, "Finding Complex Cycles through a Set of Nodes," in *Global Telecommunications Conference (GLOBECOM 2011), 2011 IEEE*, 2011.
[7] M. Maekawa, "An algorithm for mutual exclusion in decentralized systems," *ACM Transactions on Computer Systems (TOCS),* vol. 3, no. 2, pp. 145-159, 1985.
[8] W.-S. Luk and T.-T. Wong, "Two new quorum based algorithms for distributed mutual exclusion," in *Distributed Computing Systems, Proceedings of the 17th International Conference on*, 1997.